\documentclass[twocolumn,twoside,showkeys,preprintnumbers,amsmath,amssymb,secnumarabic, nofootinbib, nobibnotes,epl,floatfix]{revtex4}

\usepackage{epsfig}
\usepackage{graphicx}
\usepackage{fancyhdr}
\usepackage{pslatex}
\usepackage{longtable}

\newcommand{\be}{\begin{equation}}
\newcommand{\ee}{\end{equation}}
\newcommand{\bea}{\begin{eqnarray}}
\newcommand{\eea}{\end{eqnarray}}

\usepackage[latin2]{inputenc}
\usepackage[T1]{fontenc}

\begin{document}
\title{Introducing the  {\it q-Theil} index}

\author{M. Ausloos}%
\email{marcel.ausloos@ulg.ac.be}
\affiliation{GRAPES, ULG, B5a, Sart Tilman, B-4000 Liege, Euroland  }
\author{J. Miśkiewicz}%
\email{jamis@ift.uni.wroc.pl}
\affiliation{Institute of Theoretical Physics, Wrocław University, pl. M. Borna 9, 50-204 Wrocław, Poland }


\date{Nov. 30, 2008 }

\begin{abstract}

Starting from the idea of Tsallis on non-extensive statistical mechanics and the {\it q-entropy} 
notion, we recall the Theil index $Th$ and transform it into the $Th_q$ index. Both indices can be used to map onto themselves any time series in a non linear way. We develop an application of the $Th_q$ to
the GDP evolution of 20 rich countries in the time interval [1950 - 2003]  and search for a proof of globalization of their economies. First we  calculate the distances between the ``new'' time series  and to their mean, from which such data simple networks are constructed. We emphasize that it is useful to, and we do, take into account  different time ``parameters'': (i) the moving average time window for the raw time series to calculate the $Th_q$ index; (ii)  the moving average time window for calculating the time series distances; (iii) a correlation time lag. This allows us to deduce optimal conditions to measure the features of the network, i.e.   the appearance  in 1970 of a globalization process in the economy of such countries and the present beginning of deviations. The $q$ value hereby used is that which measures the overall data distribution and is equal to 1.8125.

 \end{abstract}



\thispagestyle{fancy}

\setcounter{page}{1}

\maketitle
 
\section{Introduction}

Since the fundamental work of Boltzmann \cite{boltz} the entropy concept has been developed and applied to a range of subjects going from elementary termodynamics and statistical mechanics through quantum physics e.g. \cite{PhysRevA.63.022105} information theory \cite{shannon} up to applications in biology e.g. \cite{biol,biol2,entropy} and economy \cite{ss5,ekon2,ekon3}. Recently Tsallis and many others in his path have shaken up the usual considerations on the entropy concept, in particular within Shannon information theory.

In fact, complex nonequilibrium systems can be often described by a superstatistics, which result of a superposition of two statistics associated with two different time scales \cite{ss1,ss2,ss3,ss4,ss5,ss6}. The methods of extracting superstatistics parameters from time series are discussed in \cite{beck:056133}. In that line of thought, a special attention can be paid to the {\it entropy of a time series}. 

On the other hand, the Theil \cite{theilbio} index is often used in economy and finance. It is defined through

\be
Th(x;N) = \frac{1}{N} {\sum_{i=t}^{N} \left(\frac{x_i}{\langle x_i \rangle}  \ln \frac{x_i}{\langle x_i \rangle}\right)}
\label{eq:defTh}
\ee
where the average  $\langle x  \rangle$ is made over the ensemble of points $N$ of the population of size $N$.
It looks like the Shannon entropy but was invented to consider the event values themselves, in particular the income $x_i$ of agent $i$ in a population of $N$ agents,  rather than their probability of occurrence. One peculiarity is that it measures the individual's share of income relative to the mean income $\langle x_i  \rangle$ of the population. 
With reference to information theory, Theil's measure is a 
difference between its maximum entropy and its present entropy at that time.
Thus from the Theil index one can look at correlations between data sets, distances, hierarchies, and other usual features, through various techniques of data analysis, like those resulting after network constructions. 

An interesting development is to consider that the $x_i$ quantity in Eq.(\ref{eq:defTh}) is time dependent. Thus one can generalize the Theil index in order to remap in a nonlinear way a time series $x(t)$ into a $Th(t)$, as done in Sect. 2 which recalls considerations outlined in  \cite{3}.
Moreover in the spirit of non-extensive statistical entropy, following Tsallis considerations, it can be imagined to propose the {\it q-Theil} index, as done in Sect. 2. 
The first application is here below made to macroeconomy time series, in particular to the GDP of the richest countries. Following up on studies of correlations between GDPs of rich countries  \cite{3,1,JMMA,MA08,5,6,gligor-2008,ausloos-2008,9},
we have analyzed web-downloaded data on GDP, used as individual wealth signatures of a country economical state (``status''). 
We have calculated the fluctuations of the GDP and looked for correlations, and ``distances'', as reported  in Sect. 3. 

Usually, a system is represented by a network, nodes being scalar agents, here the countries, while links are weights, i.e. here measures of distances between two $Th(t)$ representing GDP fluctuation correlations between two countries. 
In order to extract structures from the networks, we have averaged the time correlations in different windows. This allows more robustness in the subsequent networks properties and reveals  evolving {\it statistical  distances}. In line with our previous works \cite{1,3,JMMA,MA08} we have examined three different network constructions.
A discussion on economy globalization follows with a conclusion in Sect. 4. It is found that such a measure of {\it collective habits} does fit the usual expectations defined by politicians or economists,  i.e. common factors are to be searched for.

\section{Theil index and Tsallis entropy } 

The original definition of the Theil index, see Eq.(\ref{eq:defTh}), allows for a peculiar mapping of a 1-D data set,  like a time series, i.e. The Theil index nonlinearly maps the ``original'' time series $A(t)$ into a new one through

\be
Th_A(t,T_1) = \frac{1}{T_1} {\sum_{i=t}^{t+T_1} \left(\frac{A_i}{\langle A \rangle_{(t,T_1)}}  \ln \frac{A_i}{\langle A \rangle_{(t,T_1)}}\right)}
\label{eq:th}
\ee
where the average  $\langle A \rangle_{(t,T_1)} $ is made over the ensemble of points $j$ in a time window of size $T_1$, placed between $t$ and $t+T_1$:
\be
{\langle A \rangle_{(t,T_1)}} =\frac{1}{T_1} {\sum_{j=t}^{t+T_1} A_j}.
\label{eq:mean}
\ee
Thus the Theil index is calculated for the interval $[t,t+T_1]$. Applications of the Theil index notion can be found in other papers \cite{3,MA08}, where the Theil index was applied to measure the economy globalization process.

In order to connect with Tsallis non extensive statistics we introduce the   {\it q-Theil}  index  
$Th_{q}^A$ for a time series  $A(t)$ 
\be
Th_{q}^A(t,T_1) =
\frac{1- \sum_{i=t}^{t+T_1} \left(( {A_i}/{\langle A_i \rangle_{(t,T_1)}})^{q} \right)}{q-1}
\label{eq:thq}
\ee
in the interval $[t,t+T_1]$. Eq.(\ref{eq:thq}) corresponds to Eq.(\ref{eq:th}) when $q \rightarrow 1$.

In order to compare  time series,  their distance can be immediately introduced. Moreover the mean and standard deviations  (std) of an ensemble of such distances can be used in further considerations.
The distance between two  time series (here the Theil-mapped time series) is hereby defined as the absolute value of the difference between mean values in the interval $[t,t+T_2]$. Moreover the elements of the time series can be taken at equal times or with the time lag $\tau$,  a possibility which we take also into consideration for generality purposes. Thus we define
\be
d_{Th_q}(A,B)_{(t,T_1,T_2,\tau)} = \vert \langle Th_{q}^A(t,T_1) - Th_{q}^B (t+\tau,T_1)  \rangle_{(t,T_2)} \vert .
\label{eq:thq_d}
\ee 

In Eq.(\ref{eq:thq_d}) the mean value denoted by brackets, $\langle ... \rangle $, is defined as in Eq.(\ref{eq:mean}).

As a result we have three different time parameters:
\begin{enumerate}
\item the  $T_1$ time window while calculating the $Th_{q}$ index,
\item the time lag $\tau$, and
\item the correlation window $T_2$.
\end{enumerate}
Note: in the analysis both time windows ($Th_{q}$ and correlation) are used congruently so the the total size of the time window is equal to the sum of the $Th_{q}$ and correlation time windows. Therefore the number of the generated networks is equal to the time series length minus the total time window size.

\section{Macroeconomy index input and network construction}

\subsection{GDP data}
\label{sec:data}
GDP data sets of most rich OECD countries were used, i.e. 20 countries  Austria, Belgium, Canada, Denmark, Finland, France, Greece, Ireland, Italy, Japan, the Netherlands, Norway, Portugal, Spain, Sweden, Switzerland, Turkey, U.K., U.S.A, and Germany, allowing for a linear superposition of the data before the reunification in 1991 in the latter case; an All country is also invented as in previous works \cite{1,3,JMMA,MA08}.\footnote{The set deviates somewhat from previous works \cite{3,JMMA,MA08} since there is neither Iceland nor Luxembourg but there is Turkey in the present paper.}  Thus $N=$ 21.
The data starts in 1950 and finish in 2003, so there are 54 data points in every time series. 

\subsection{Networks}
The distance matrices obtained from Eq.(\ref{eq:thq_d})  are analysed by constructing three network structures and analysing statistical properties of the distances between nodes. The following networks are considered: unidirectional minimal length path (UMLP), bidirectional minimal length path (BMLP) and locally minimal spanning tree (LMST). The algorithms generating the mentioned networks are:
\begin{description}
\item[UMLP]
The network begins with an arbitrary chosen country, -- here the All country, then the closest neighbouring country is attached and become the end of the network. The next country closest to the end of the network is searched and attached. The process continued until all countries are attached to the network.
\item[BMLP]
The network begins with the pair of countries with the smallest distance between them. Then the country closest to the ends of the network are searched and those with shorter distance attached to the appropriate end. The algorithm is continued until all countries become nodes of the network.
\item[LMST]
The root of the network is the pair of closest neighbouring countries. Then the country closest to any node is searched and attached. The algorithm is continued until all countries are attached to the network.

 \end{description}

Notice that  in the UMLP construction,  All is at the begining of the chain, while in the other two constructions, All is treated as a ''normal'' country. The BMLP and LMST network seeds are the appropriate pairs of the closest  countries according to the appropriate distance matrix.
The first two networks are linear, and essentially robust against a ``perturbation'', like removing or adding a country or in the case of a regrettable mathematical error, since they are based on a measure relative to a statistical mean, while the LMST is obvioulsy a tree, rather compact when only 21 data points, thus with very few branching levels, are involved in the construction. It is known that such a tree is far from robust.

\section{Results}

First let us report that a $q$ value of the considered data set must be given.   It could be let as a free parameter and one could find some optimal value according to some criterion, or a few criteria. Here below, i.e. for the GDP time series in 1950-2003 for the countries defined in Sec.\ref{sec:data} we have calculated the $q$ value  for the following considerations by the maximum likelihood estimator \cite{q-calc}, as for Tsallis entropy, and found   $q=1.8315$, hereby used to calculate distances through (Eq.(\ref{eq:thq}) ) and Eq.(\ref{eq:thq_d}). It is fair to recall that Borges in  \cite{qGDP} calculated the (Tsallis entropy) $q$ value  for GDP of USA, Brasil, Germany and UK. He found a $q$ value varying from 1.4 (UK) up to 2.1 (Brasil), and or USA, $ q$=1.7.

\subsection{{\it q-Theil}  distance statistics   }

In our analysis UMLP, BMLP and LMST networks were constructed for all time windows ranging from $T_{1}=5$ yrs , $T_2=1$ y  moving along the time axis by a one year step. Eleven time lag values were considered: $\tau \in [0,1, \ldots ,10]$. The $T_1$, $T_2$ and $\tau$ parameters statisfy the inequality $T_2+ T_3 + \tau \leq 54$ yrs, so the number of generated networks ($Net$) depends on the time window sizes and is equal to $N_{Net}=54-T_1-T_2-\tau$, for a given triplet, - times 3, due to the type of network considered. In total this is a huge number of networks. Therefore some cases are to be extracted for the present report.\footnote{All cases are available from the authors upon request.} Different presentations can be made, in a three dimensional time coordinate space. We propose a vizualisation of the data through a spectrogram method, using for the $x$ and $y$ axis the time window $T_2$ and $T_1$ respectively for a given $\tau$. The data values are represented by a colored pixel in a convenient order.\footnote{The results are presented in grey tones, but online figures are available in color.}  The results of calculations of the mean but also  values of the corresponding standard deviations  are here below presented.

\begin{table*}
\begin{tabular}{|l|l|l|l|l|l|l|l|l|l|l|l|l|l|l|l|l|l|l|} \hline
\multicolumn{2}{|l|}{$q=1.8315$}   & \multicolumn{8}{c|}{mean}& \multicolumn{8}{c|}{std}\\ \cline{1-18}

Network&$\tau$ (y)& max & $T_{1}$ & $T_2$ & $N_{Net}$&min & $T_{1}$ & $T_2$ & $N_{Net}$&  max & $T_{1}$ & $T_2$& $N_{Net}$ &min& $T_{1}$ & $T_2$ & $N_{Net}$\\ \hline
&  0   & 1.110 & 52 & 1 & 1 & $18 \cdot 10^{-4} $& 5 & 48 & 1 & 1.549 & 52 & 1 & 1 &$29 \cdot 10^{-4} $& 5 & 48 & 1\\ \cline{2-18}
UMLP & 5  & 1.956 & 43 & 5 & 1 &$37 \cdot 10^{-4} $& 5 & 37 & 7&2.414 & 43 & 5 & 1 &$45 \cdot 10^{-4} $& 5 & 43 & 1\\ \cline{2-18}
& 10 & 2.984 & 41 & 1 & 2 & $61 \cdot 10^{-4} $& 5 & 35 & 4 &3.151 & 41 & 2 & 1 &$72 \cdot 10^{-4} $& 5 & 38 & 1  \\ \hline
& 0 & 0.8151 & 52 & 1 & 1 &$15 \cdot 10^{-4} $& 5 & 47 & 2&2.7829 & 51 & 2& 17 &$24 \cdot 10^{-4} $&5 &48&1  \\ \cline{2-18}
BMLP & 5 & 2.0014 & 46 & 2 & 1 & $28 \cdot 10^{-5} $& 5 & 41 & 3&2.0473 & 46 & 2 & 5 &$41 \cdot 10^{-4} $& 5&41&3  \\ \cline{2-18}
& 10 & 2.7829 & 40 & 2 & 2 &$56 \cdot 10^{-4} $& 5 & 37 & 2&3.0620 & 40 & 2 & 2 &$69 \cdot 10^{-4} $& 5 &37&2\\ \hline
& 0  & 0.8151 & 52 & 1 & 1 &$15 \cdot 10^{-4} $&5&47 &2&0.7671 & 51 & 2 & 1 &$24 \cdot 10^{-4} $& 5 &48&1\\ \cline{2-18}
LMST & 5 & 1.9609 & 45 & 2& 2 &$28 \cdot 10^{-4} $&5&43  &1&1.8710 & 45 & 2 & 2 &$43 \cdot 10^{-4} $& 5&39&5 \\ \cline{2-18}
& 10 & 3.2012 & 40 & 2 & 2 &$52 \cdot 10^{-4} $&5&38 &1&3.2334 & 41 & 2 & 1 & $68 \cdot 10^{-4} $ & 5 &38&1\\ \hline
\end{tabular}
\caption{   The maximum mean distance, the minimum mean distance, the maximum and minimum standard deviations  resulting for each type of network, for a few characteristics $\tau$ values are given
when  $Th_{q}$  is calculated  for $q=1.8315$.
Recall that the ``mean''  is that of the distances between nodes on the indicated network in the ensemble of networks generated for the given time windows. The values of the averaging windows $T_1$ and $T_2$ when this maximum (minimum) occurs are indicated; the corresponding number of networks ($N_{Net}$) used for the statistics is also indicated.}
\label{tab:q-Theil}
\end{table*}

\begin{figure}
 \centering
\includegraphics[bb=50 50 266 201]{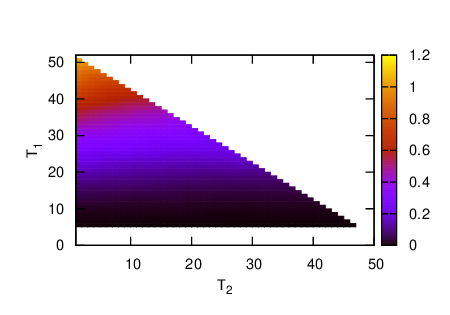}
\caption{Mean distance between countries  as obtained in the case of the $Th_{\mathbf{q}}$ mapping and the UMLP network construction. The distance is averaged over the network links and the time. The values of $T_1$ and $T_2$ are given on the vertical and horizontal axis respectively. Time lag value: $\tau=0$ y. The color scale  in use is indicated.  A few numerical values are given in Table \ref{tab:q-Theil}.}
\label{fig:tsallis_umlp_m0}
\end{figure}

\begin{figure}
 \centering
\includegraphics[bb=50 50 266 201]{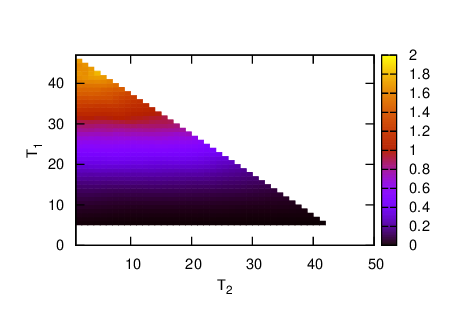}
 
\caption{Mean distance between countries deduced from $Th_{\mathbf{q}}$    for UMLP networks. The distance is averaged over the network links and the time.
The size of $T_1$ and $T_2$ are presented on the vertical and horizontal axis respectively. Time lag    $\tau=5$ yrs. The color scale  in use is indicated.  A few numerical values are given in   Table \ref{tab:q-Theil}.}\label{fig:tsallis_umlp_m5}
\end{figure}

\begin{figure}
 \centering
\includegraphics[bb=50 50 266 201]{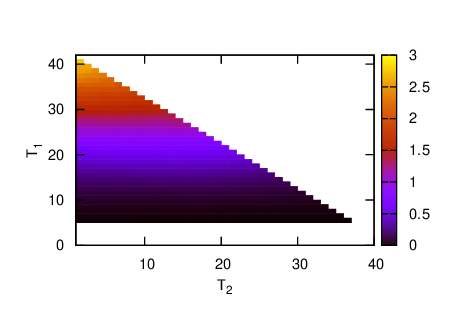}
 \caption{Mean distance between countries deduced from $Th_{\mathbf{q}}$   for UMLP networks. The distance is averaged over the network links and the time. The size of $T_1$ and $T_2$ are presented on the vertical and horizontal axis respectively. Time lag    $\tau=10$ yrs. The color scale  in use is indicated.  A few numerical values are given in   Table \ref{tab:q-Theil}.}
\label{fig:tsallis_umlp_m10}
\end{figure}

\begin{figure}
 \centering
\includegraphics[bb=50 50 266 201]{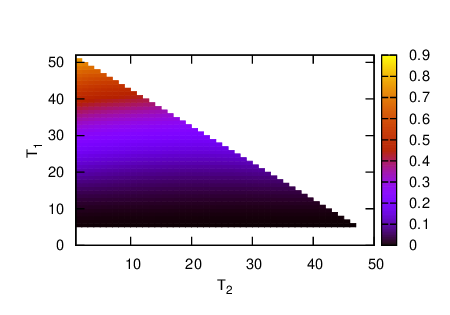}
\caption{Mean distance between countries deduced from $Th_{\mathbf{q}}$  for  BMLP networks. The distance is averaged over the network links and the time. The size of $T_1$ and $T_2$ are presented on the vertical and horizontal axis respectively. Time lag  $\tau=0$ y. The color scale  in use is indicated.  A few numerical values are given in   Table \ref{tab:q-Theil}.}
 \label{fig:tsallis_bmlp_m0}
\end{figure}

\begin{figure}
 \centering
\includegraphics[bb=50 50 266 201]{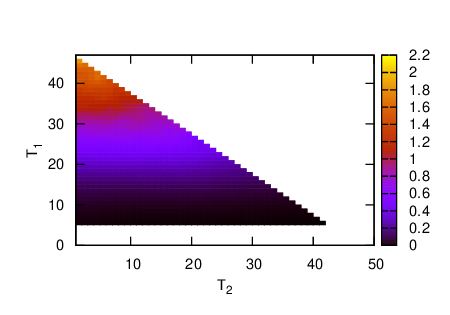}
\caption{Mean distance between countries deduced from $Th_{\mathbf{q}}$  for BMLP networks. The distance is averaged over the network links and the time. The size of $T_1$ and $T_2$ are presented on the vertical and horizontal axis respectively. Time lag  $\tau=5$ yrs. The color scale  in use is indicated.  A few numerical values are given in Table \ref{tab:q-Theil}.} \label{fig:tsallis_bmlp_m5}
\end{figure}

\begin{figure}
 \centering
\includegraphics[bb=50 50 266 201]{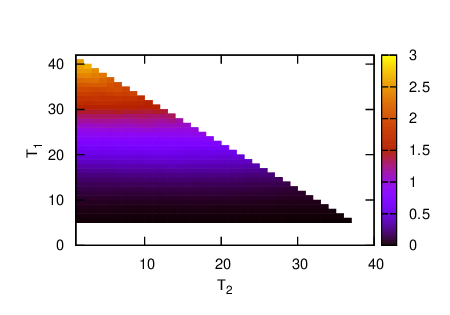}
\caption{Mean distance between countries  deduced from $Th_{\mathbf{q}}$    for  BMLP networks. The distance is averaged over the network links and the time. The size of $T_1$ and $T_2$ are presented on the vertical and horizontal axis respectively. Time lag    $\tau=10$ yrs. The color scale  in use is indicated.  A few numerical values are given in   Table \ref{tab:q-Theil}.} \label{fig:tsallis_bmlp_m10}
\end{figure}

\begin{figure}
\centering
\includegraphics[bb=50 50 266 201]{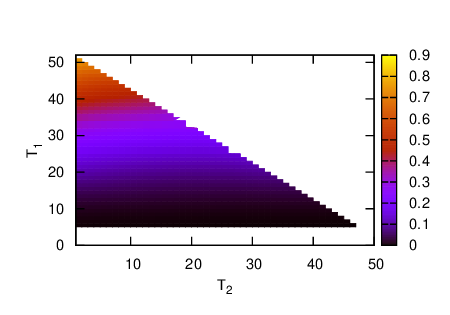}
\caption{Mean distance between countries  deduced from $Th_{\mathbf{q}}$    for LMST networks. The distance is averaged over the network links and the time. The size of $T_1$ and $T_2$ are presented on the vertical and horizontal axis respectively. Time lag  $\tau=0$ y. The color scale  in use is indicated.  A few numerical values are given in   Table \ref{tab:q-Theil}.}
 \label{fig:tsallis_lmst_m0}
\end{figure}

\begin{figure}
\centering
\includegraphics[bb=50 50 266 201]{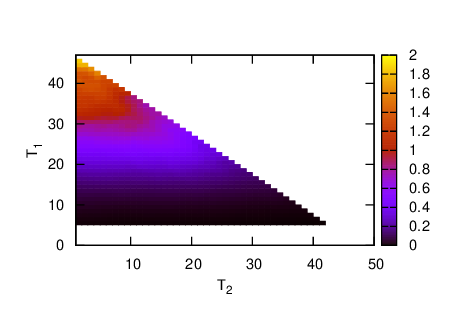}
\caption{Mean distance between countries deduced from $Th_{\mathbf{q}}$    for  LMST network. The distance is averaged over the network links and the time. The size of $T_1$ and $T_2$ are presented on the vertical and horizontal axis respectively. Time lag  $\tau=5$ yrs. The color scale  in use is indicated.  A few numerical values are given in   Table \ref{tab:q-Theil}.}
 \label{fig:tsallis_lmst_m5}
\end{figure}

\begin{figure}
\centering
\includegraphics[bb=50 50 266 201]{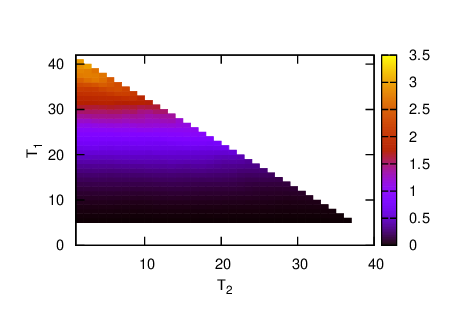}
\caption{Mean distance between countries deduced from $Th_{\mathbf{q}}$    for  LMST networks. The distance is averaged over the network links and the time. The size of $T_1$ and $T_2$ are presented on the vertical and horizontal axis respectively. Time lag  $\tau=0$ yrs. The color scale  in use is indicated.  A few numerical values are given in   Table \ref{tab:q-Theil}.}
 \label{fig:tsallis_lmst_m10}
\end{figure}

The mean value and standard deviation of the distances between nodes as a function of the $ T_1$ and $T_2$ are presented in Figs. \ref{fig:tsallis_umlp_m0} - \ref{fig:tsallis_lmst_m10} for the time lag $\tau=0,\;5,\;10$ yrs. The largest value of the mean distance, the minimum mean distance, the maximum and minimum standard deviations as a function of the time windows $T_1$, $T_2$ and time lags are presented in Table \ref{tab:q-Theil}.

It can be first generally observed that the mean distance between countries and the corresponding standard deviation are the biggest for UMLP networks and the smallest for LMST networks. It is also worth noticing that the mean distance depends on the time lag value. If the time lag is large the mean distance is large as well. The maximum of the mean distance occurs for the longest $T_1$ and the shortest $T_2$ windows sizes. The minimum mean distance is found with the oposite combination of the time windows sizes, i.e. small $T_1$ and large $T_2$. The standard deviations increase with the time lag and are the largest ones in the case of the longest considered time lag.

\begin{figure}
 \centering
 \includegraphics[bb=50 50 410 302,scale=0.5]{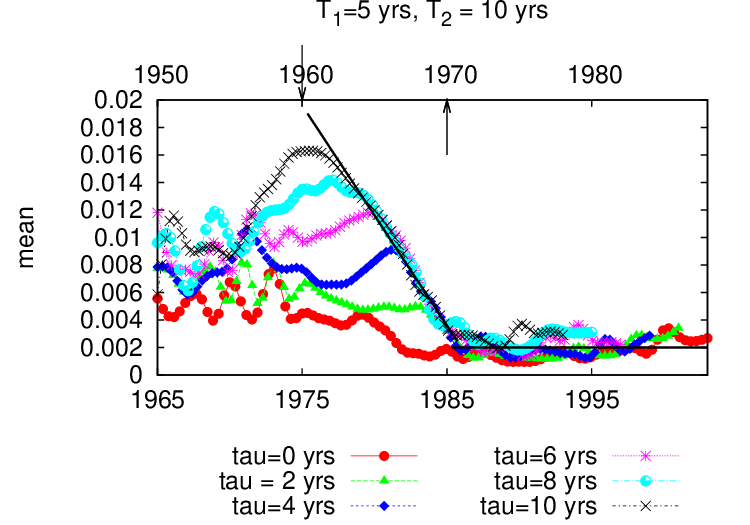}
\caption{Yearly evolution of the mean and standard deviation of the links between nodes for the UMLP network deduced from $Th_{\mathbf{q}}$    analysis of GDP countries, when  $T_1=5$ yrs, $T_2=10$ yrs  for different time lags $\tau$.
}
 \label{fig:tsallis_umlp_hist_5_10}
\end{figure}

\begin{figure}
 \centering
 \includegraphics[bb=50 50 410 302,scale=0.5]{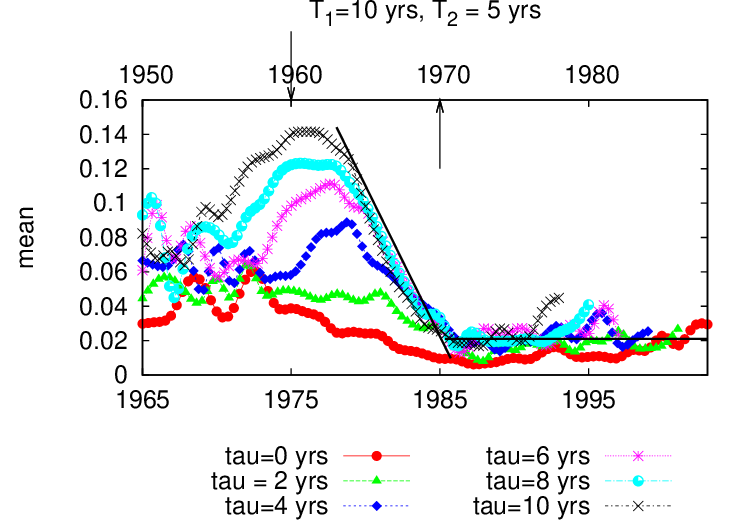}
\caption{Yearly evolution of the mean and standard deviation of the links between nodes for the UMLP network deduced from $Th_{\mathbf{q}}$    analysis of GDP countries, when  $T_1=10$ yrs, $T_2=5$  yrs  for different time lags $\tau$.}
 \label{fig:tsallis_umlp_hist_10_5}
\end{figure}

\begin{figure}
 \centering
 \includegraphics[bb=50 50 410 302,scale=0.5]{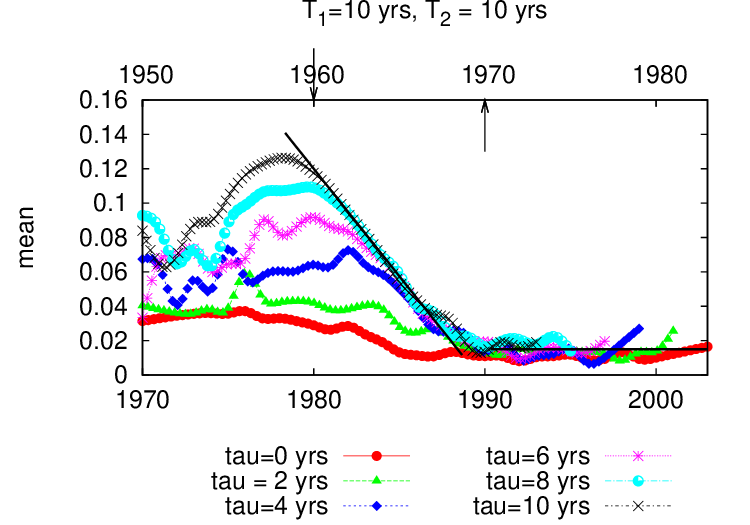}
\caption{Yearly evolution of the mean and standard deviation of the links between nodes for the UMLP network deduced from $Th_{\mathbf{q}}$    analysis of GDP countries, when  $T_1=10 $ yrs, $T_2=10$  yrs  for different time lags $\tau$.}
 \label{fig:tsallis_umlp_hist_10_10}
\end{figure}

\begin{figure}
 \centering
 \includegraphics[bb=50 50 410 302,scale=0.5]{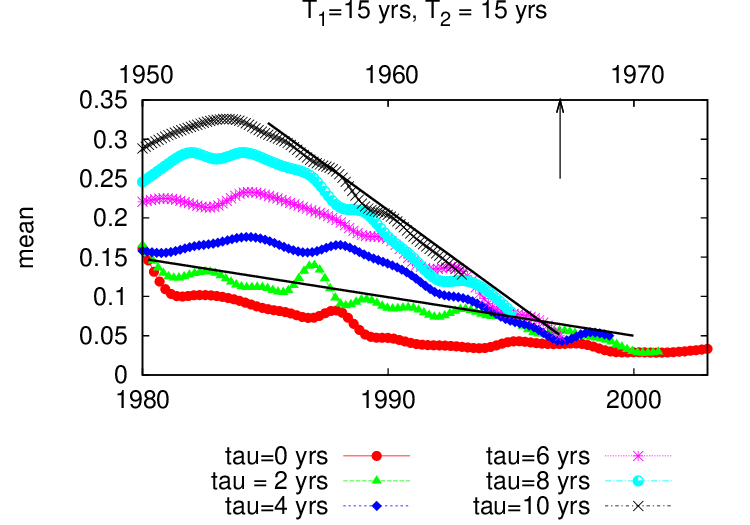}
\caption{Yearly evolution of the mean and standard deviation of the links between nodes for the UMLP network deduced from $Th_{\mathbf{q}}$   analysis of GDP countries, when  $T_1=15 $ yrs, $T_2=15$  yrs  for different time lags $\tau$.
}
\label{fig:tsallis_umlp_hist_15_15}
\end{figure}

\begin{figure}
 \centering
 \includegraphics[bb=50 50 410 302,scale=0.5]{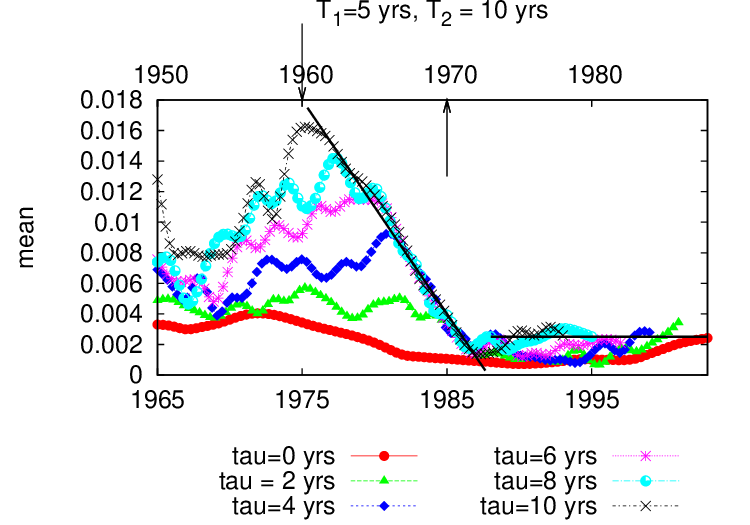}
\caption{Yearly evolution of the mean and standard deviation of the links between nodes for the BMLP network deduced from $Th_{\mathbf{q}}$    analysis of GDP countries, when  $T_1=5 $ yrs, $T_2=10$ yrs  for different time lags $\tau$.}
 \label{fig:tsallis_bmlp_hist_5_10}
\end{figure}

\begin{figure}
 \centering
 \includegraphics[bb=50 50 410 302,scale=0.5]{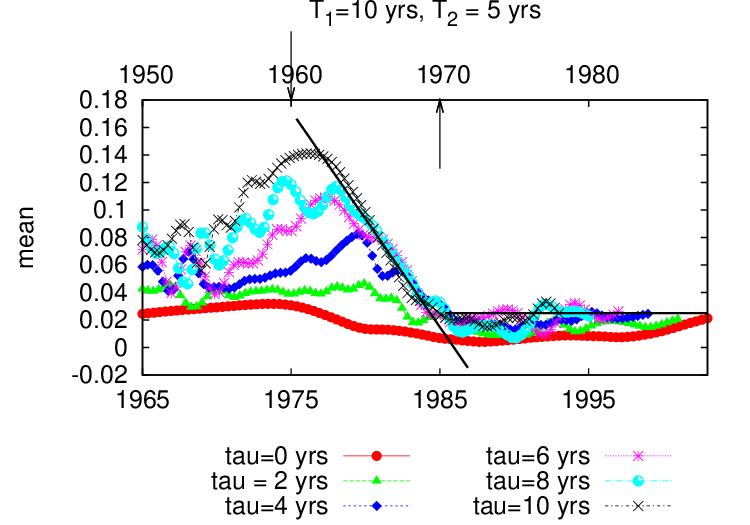}
 \caption{Yearly evolution of the mean and standard deviation of the links between nodes for the BMLP network deduced from $Th_{\mathbf{q}}$   analysis of GDP countries, when  $T_1=10$ yrs, $T_2=5$ yrs   for different time lags $\tau$.}
\label{fig:tsallis_bmlp_hist_10_5}
\end{figure}

\begin{figure}
 \centering
 \includegraphics[bb=50 50 410 302,scale=0.5]{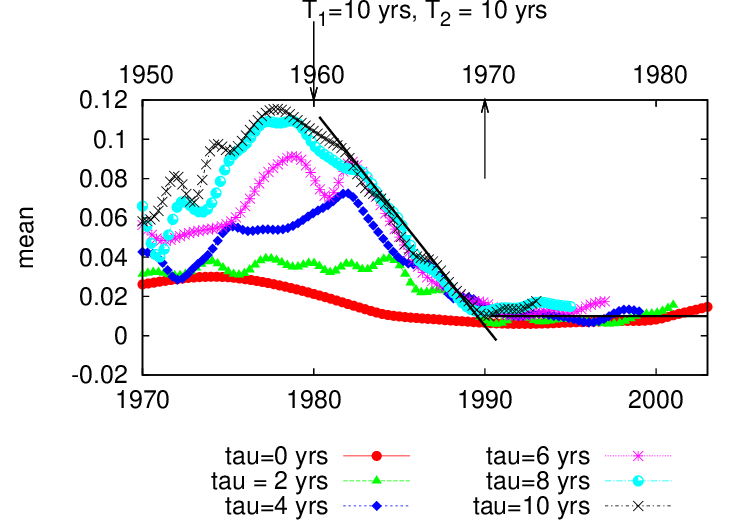}
 \caption{Yearly evolution of the mean and standard deviation of the links between nodes for the BMLP network deduced from $Th_{\mathbf{q}}$   analysis of GDP countries, when  $T_1=10$ yrs, $T_2=10$ yrs   for different time lags $\tau$.}
\label{fig:tsallis_bmlp_hist_10_10}
\end{figure}

\begin{figure}
 \centering
 \includegraphics[bb=50 50 410 302,scale=0.5]{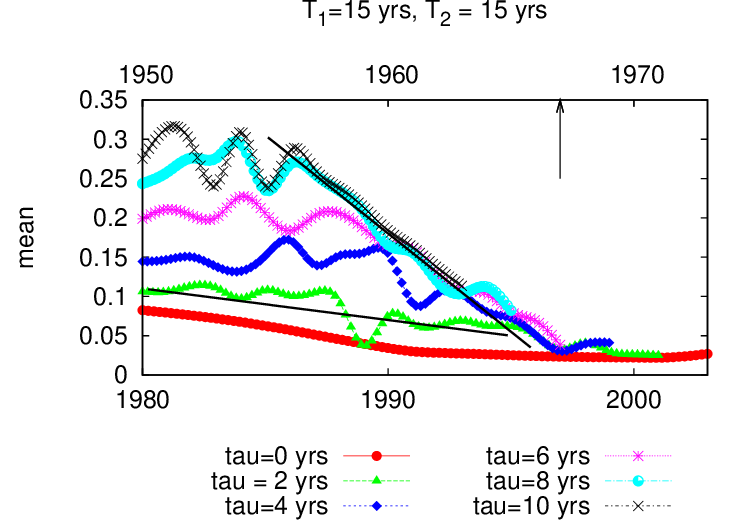}
 \caption{Yearly evolution of the mean and standard deviation of the links between nodes for the BMLP network deduced from $Th_{\mathbf{q}}$    analysis of GDP countries, when  $T_1=15$ yrs, $T_2=15 $ yrs  for different time lags $\tau$.}
\label{fig:tsallis_bmlp_hist_15_15}
\end{figure}

\begin{figure}
 \centering
 \includegraphics[bb=50 50 410 302,scale=0.5]{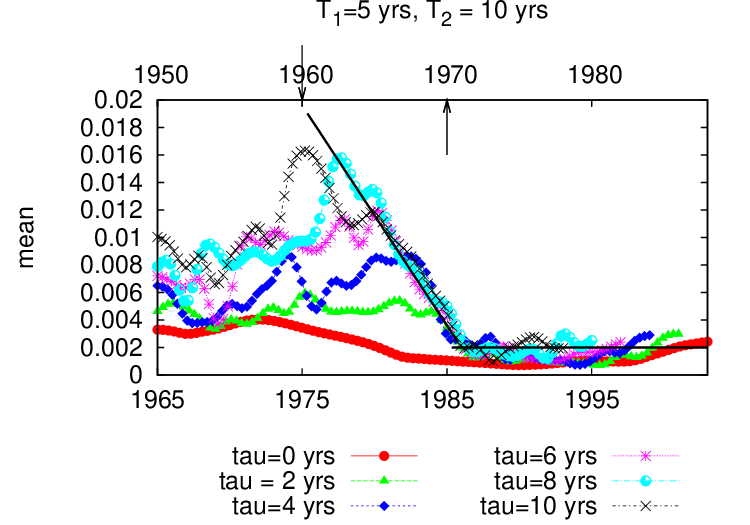}
 \caption{Yearly evolution of the mean and standard deviation of the links between nodes for the LMST network deduced from $Th_{\mathbf{q}}$    analysis of GDP countries, when  $T_1=5$ yrs, $T_2=10$ yrs  for different time lags $\tau$.
}
\label{fig:tsallis_lmst_hist_5_10}
\end{figure}

\begin{figure}
 \centering
 \includegraphics[bb=50 50 410 302,scale=0.5]{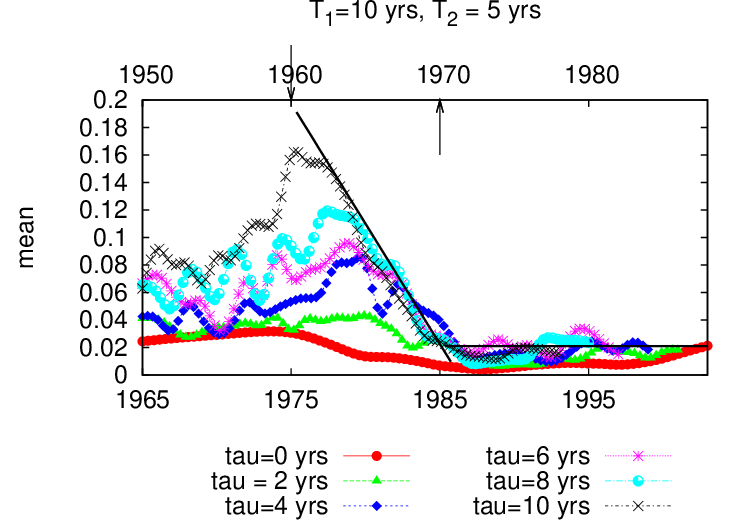}
 \caption{Yearly evolution of the mean and standard deviation of the links between nodes for the LMST network deduced from $Th_{\mathbf{q}}$   analysis of GDP countries, when  $T_1=10$  yrs, $T_2=5$ yrs  for different time lags $\tau$.
}
\label{fig:tsallis_lmst_hist_10_5}
\end{figure}

\begin{figure}
 \centering
 \includegraphics[bb=50 50 410 302,scale=0.5]{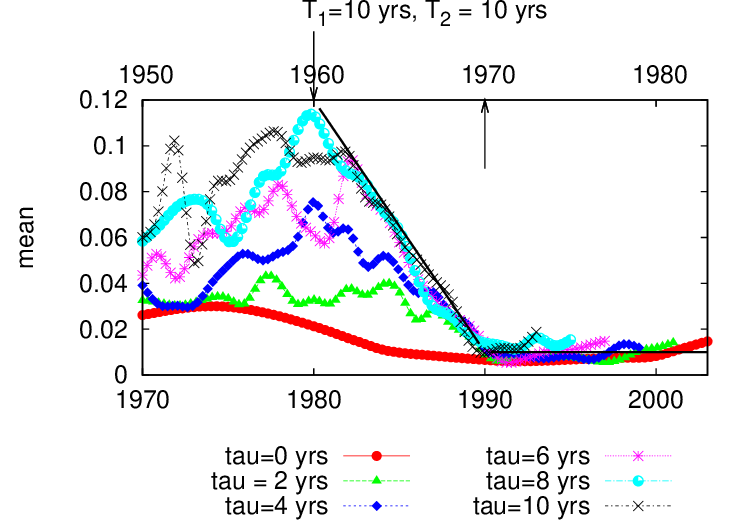}
 \caption{Yearly evolution of the mean and standard deviation of the links between nodes for the LMST network deduced from $Th_{\mathbf{q}}$    analysis of GDP countries, when  $T_1=10$ yrs, $T_2=10 $ yrs  for different time lags $\tau$.}
\label{fig:tsallis_lmst_hist_10_10}
\end{figure}

\begin{figure}
 \centering
 \includegraphics[bb=50 50 410 302,scale=0.5]{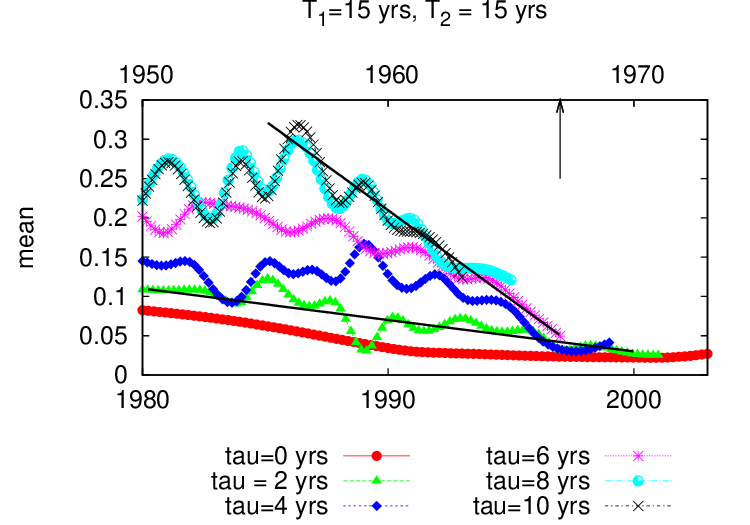}
 \caption{Yearly evolution of the mean and standard deviation of the links between nodes for the LMST network deduced from $Th_{\mathbf{q}}$   analysis of GDP countries, when  $T_1=15 $ yrs, $T_2=15$  yrs  for different time lags $\tau$.}
\label{fig:tsallis_lmst_hist_15_15}
\end{figure}

\subsection{{\it q-Theil} network evolution }

For further discussion the  following  time window  sizes were chosen, i.e. $(T_{1}=5$ yrs, $T_2=10$ yrs$)$, $(T_{1}=10$ yrs, $T_2=5$ yrs$)$, $(T_{1}=10$ yrs, $ T_2=10$ yrs$)$, $(T_{1}=15$ yrs, $ T_2=15$ yrs$)$, for the three time lags $\tau = 0, 5, 10$ yrs. The evolutions of mean distance between countries and the corresponding standard deviations for these chosen time windows sizes are presented in Figs. \ref{fig:tsallis_umlp_hist_5_10}-\ref{fig:tsallis_lmst_hist_15_15}.  Arrows and straight lines indicate remarkable features.

The general observations to be made at this stage are the following
\begin{itemize}
\item 
In all considered networks (UMLP, BMLP and LMST) and for all window sizes three types of evolution can be distinguished: increase, decrease and relatively stable mean distance between countries. 
\item These three types of evolution are better seen for long lag time ($\tau > 5 $ yrs). Therefore the lag time seems to be crucial in any analysis and discussion of the globalization process. This might suggest  that  some countries play a role of leaders while other follow their way. 
\item It is worth noticing that for the very long lag time $\tau=10$ yrs and time windows [$(T_{1}=5 $ yrs, $T_2=10$ yrs$)$, $(T_{1}=10$ yrs, $ T_2=5 $yrs$)$, $(T_{1}=10 $yrs, $ T_=10 $yrs$)$] the maximum of the mean distance occurs about 1960,  and 
\item since then the size of the network(s) is fast decreasing over a decade up to 1970 and 
\item thereafter remains small and relatively stable up to 2000 or so
\item when the mean size seems to reincrease.
\end{itemize}

\section{Conclusions}

In conclusion, the most interesting results of this analysis are
\begin{itemize}
\item The analysis shows the existence of a globalization process since  1960 till 1970 and its stabilisation thereafter, followed by a destabilisation after 2000 as observed in the decrease of the network size. 
\item The observation of the globalization process does not depend on the type of network constructed.
\item The mean distance between countries and the corresponding std are the largest for the UMLP networks and the smallest for the corresponding LMST networks.
\item With increasing time lag the Theil mapping window size $T_1$ at which the maximum of the network size is found is always decreasing.
 \item The globalization process is better seen if the lag time is greater than 5 yrs, - which might be
considered as the time needed for some synchronization process, but is  also in fact commensurate with most government life times and election time intervals. These conjectures suggest further investigations.
\item Even though for large time lags the mean values are large,   the globalization evolution is  the same as for short time lags (greater than 5 yrs);   thus a large time lag magnifies the globalization process feature which is easilier to observe then.
\end{itemize}

Of course much more work is in order to connect the above to some non extensive  thermostatistics ideas.  To search for  a robust  (``optimal'')  $q$ value and  the significance of the {\it q-Theil } index are open questions. Finally let us stress the interest of  studying
graphs,  in particular to derive weighted networks such as in this paper, in order to have some comparative data organisation coherence.

\subsection{Acknowldgements}

Thanks to the organizers of NEXT2008 in Foz do Iguacu, Parana, in particular Luis C. Malacarne and Reino S. Mendes, plus kind thanks to Thais  Pedreira for their welcome and careful attention to MA needs at the meeting. Let C. Tsallis be congratulated for his great insight,  what he brought to modern equilibrium thermostatistics and  for suggesting  MA participation at such a meeting.

\bibliographystyle{apsrev}

\bibliography{NEXTheilTsallis}

\begin{thebibliography}{27}
\expandafter\ifx\csname natexlab\endcsname\relax\def\natexlab#1{#1}\fi
\expandafter\ifx\csname bibnamefont\endcsname\relax
  \def\bibnamefont#1{#1}\fi
\expandafter\ifx\csname bibfnamefont\endcsname\relax
  \def\bibfnamefont#1{#1}\fi
\expandafter\ifx\csname citenamefont\endcsname\relax
  \def\citenamefont#1{#1}\fi
\expandafter\ifx\csname url\endcsname\relax
  \def\url#1{\texttt{#1}}\fi
\expandafter\ifx\csname urlprefix\endcsname\relax\def\urlprefix{URL }\fi
\providecommand{\bibinfo}[2]{#2}
\providecommand{\eprint}[2][]{\url{#2}}

\bibitem[{\citenamefont{Boltzmann}(1895/98)}]{boltz}
\bibinfo{author}{\bibfnamefont{L.}~\bibnamefont{Boltzmann}},
  \emph{\bibinfo{title}{Vorlesungen {\"u}ber Gastheorie: 2 Volumes}}
  (\bibinfo{publisher}{Leipzig}, \bibinfo{year}{1895/98}).

\bibitem[{\citenamefont{Gheorghiu-Svirschevski}(2001)}]{PhysRevA.63.022105}
\bibinfo{author}{\bibfnamefont{S.}~\bibnamefont{Gheorghiu-Svirschevski}},
  \bibinfo{journal}{Phys. Rev. A} \textbf{\bibinfo{volume}{63}},
  \bibinfo{pages}{022105} (\bibinfo{year}{2001}).

\bibitem[{\citenamefont{Shannon}(1950)}]{shannon}
\bibinfo{author}{\bibfnamefont{C.~E.} \bibnamefont{Shannon}},
  \bibinfo{journal}{The Bell System Technical Journal}
  \textbf{\bibinfo{volume}{30}}, \bibinfo{pages}{50} (\bibinfo{year}{1950}).

\bibitem[{\citenamefont{{Ruiz de la Torre} et~al.}(2000)\citenamefont{{Ruiz de
  la Torre}, Velarde, and Manteca}}]{biol}
\bibinfo{author}{\bibfnamefont{J.~L.} \bibnamefont{{Ruiz de la Torre}}},
  \bibinfo{author}{\bibfnamefont{A.}~\bibnamefont{Velarde}}, \bibnamefont{and}
  \bibinfo{author}{\bibfnamefont{X.}~\bibnamefont{Manteca}},
  \bibinfo{journal}{Animal Behaviour} \textbf{\bibinfo{volume}{59}},
  \bibinfo{pages}{269} (\bibinfo{year}{2000}).

\bibitem[{\citenamefont{Brooks and Wiley}(1988)}]{biol2}
\bibinfo{author}{\bibfnamefont{D.~R.} \bibnamefont{Brooks}} \bibnamefont{and}
  \bibinfo{author}{\bibfnamefont{E.~O.} \bibnamefont{Wiley}},
  \emph{\bibinfo{title}{Evolution As Entropy: Toward a Unified Theory of
  Biology}} (\bibinfo{publisher}{University of Chicago Press},
  \bibinfo{year}{1988}).

\bibitem[{\citenamefont{{Murray Gell-Mann}}(2004)}]{entropy}
\bibinfo{author}{\bibfnamefont{C.~T.} \bibnamefont{{Murray Gell-Mann}}},
  \emph{\bibinfo{title}{Nonextensive Entropy: Interdisciplinary Applications}}
  (\bibinfo{publisher}{Oxford University Press}, \bibinfo{year}{2004}).

\bibitem[{\citenamefont{Ausloos and Ivanova}(2003)}]{ss5}
\bibinfo{author}{\bibfnamefont{M.}~\bibnamefont{Ausloos}} \bibnamefont{and}
  \bibinfo{author}{\bibfnamefont{K.}~\bibnamefont{Ivanova}},
  \bibinfo{journal}{Phys. Rev. E} \textbf{\bibinfo{volume}{68}},
  \bibinfo{pages}{046122} (\bibinfo{year}{2003}).

\bibitem[{\citenamefont{Duro and Esteban}(1998)}]{ekon2}
\bibinfo{author}{\bibfnamefont{J.~A.} \bibnamefont{Duro}} \bibnamefont{and}
  \bibinfo{author}{\bibfnamefont{J.}~\bibnamefont{Esteban}},
  \bibinfo{journal}{Economics Letters} \textbf{\bibinfo{volume}{60}},
  \bibinfo{pages}{269} (\bibinfo{year}{1998}).

\bibitem[{\citenamefont{James and Thomas}(2000)}]{ekon3}
\bibinfo{author}{\bibfnamefont{J.~A.} \bibnamefont{James}} \bibnamefont{and}
  \bibinfo{author}{\bibfnamefont{M.}~\bibnamefont{Thomas}},
  \bibinfo{journal}{Journal of Income Distribution}
  \textbf{\bibinfo{volume}{9}}, \bibinfo{pages}{39} (\bibinfo{year}{2000}).

\bibitem[{\citenamefont{Beck and Cohen}(2003)}]{ss1}
\bibinfo{author}{\bibfnamefont{C.}~\bibnamefont{Beck}} \bibnamefont{and}
  \bibinfo{author}{\bibfnamefont{E.~G.~D.} \bibnamefont{Cohen}},
  \bibinfo{journal}{Physica A} \textbf{\bibinfo{volume}{322}},
  \bibinfo{pages}{267} (\bibinfo{year}{2003}).

\bibitem[{\citenamefont{Mathai and Haubold}(2007)}]{ss2}
\bibinfo{author}{\bibfnamefont{A.}~\bibnamefont{Mathai}} \bibnamefont{and}
  \bibinfo{author}{\bibfnamefont{H.}~\bibnamefont{Haubold}},
  \bibinfo{journal}{Physica A} \textbf{\bibinfo{volume}{385}},
  \bibinfo{pages}{493} (\bibinfo{year}{2007}).

\bibitem[{\citenamefont{Touchette and Beck}(2005)}]{ss3}
\bibinfo{author}{\bibfnamefont{H.}~\bibnamefont{Touchette}} \bibnamefont{and}
  \bibinfo{author}{\bibfnamefont{C.}~\bibnamefont{Beck}},
  \bibinfo{journal}{Phys. Rev. E} \textbf{\bibinfo{volume}{71}},
  \bibinfo{pages}{016131} (\bibinfo{year}{2005}).

\bibitem[{\citenamefont{Tsallis and Souza}(2003)}]{ss4}
\bibinfo{author}{\bibfnamefont{C.}~\bibnamefont{Tsallis}} \bibnamefont{and}
  \bibinfo{author}{\bibfnamefont{A.~M.~C.} \bibnamefont{Souza}},
  \bibinfo{journal}{Phys. Rev. E} \textbf{\bibinfo{volume}{67}},
  \bibinfo{pages}{026106} (\bibinfo{year}{2003}).

\bibitem[{\citenamefont{Bouchaud and Potters}(2003)}]{ss6}
\bibinfo{author}{\bibfnamefont{J.-P.} \bibnamefont{Bouchaud}} \bibnamefont{and}
  \bibinfo{author}{\bibfnamefont{M.}~\bibnamefont{Potters}},
  \emph{\bibinfo{title}{Theory of Financial Risk and Derivative Pricing}}
  (\bibinfo{publisher}{Cambridge University Press}, \bibinfo{year}{2003}).

\bibitem[{\citenamefont{Beck et~al.}(2005)\citenamefont{Beck, Cohen, and
  Swinney}}]{beck:056133}
\bibinfo{author}{\bibfnamefont{C.}~\bibnamefont{Beck}},
  \bibinfo{author}{\bibfnamefont{E.~G.~D.} \bibnamefont{Cohen}},
  \bibnamefont{and} \bibinfo{author}{\bibfnamefont{H.~L.}
  \bibnamefont{Swinney}}, \bibinfo{journal}{Phys. Rev. E}
  \textbf{\bibinfo{volume}{72}}, \bibinfo{pages}{056133}
  (\bibinfo{year}{2005}).

\bibitem[{the()}]{theilbio}
\emph{\bibinfo{title}{{H. Theil was a Dutch econometrician who was born on13
  October 1924 in Amsterdam, graduated from the University of Amsterdam,
  succeeded to Jan Tinbergen at the Erasmus University Rotterdam, moved and
  taught later in Chicago and at the University of Florida. He died in 2000.}}}

\bibitem[{\citenamefont{Mi\'skiewicz}(2008)}]{3}
\bibinfo{author}{\bibfnamefont{J.}~\bibnamefont{Mi\'skiewicz}},
  \bibinfo{journal}{Physica A} \textbf{\bibinfo{volume}{387}},
  \bibinfo{pages}{6595} (\bibinfo{year}{2008}).

\bibitem[{\citenamefont{Mi\'skiewicz and Ausloos}(2005)}]{1}
\bibinfo{author}{\bibfnamefont{J.}~\bibnamefont{Mi\'skiewicz}}
  \bibnamefont{and} \bibinfo{author}{\bibfnamefont{M.}~\bibnamefont{Ausloos}},
  \bibinfo{journal}{Acta Phys. Pol. B} \textbf{\bibinfo{volume}{36}},
  \bibinfo{pages}{2477} (\bibinfo{year}{2005}).

\bibitem[{\citenamefont{Mi\'skiewicz and Ausloos}(2006)}]{JMMA}
\bibinfo{author}{\bibfnamefont{J.}~\bibnamefont{Mi\'skiewicz}}
  \bibnamefont{and} \bibinfo{author}{\bibfnamefont{M.}~\bibnamefont{Ausloos}},
  \bibinfo{journal}{Int. J. Mod. Phys. C} \textbf{\bibinfo{volume}{17}},
  \bibinfo{pages}{317} (\bibinfo{year}{2006}).

\bibitem[{\citenamefont{Mi\'skiewicz and Ausloos}(2008)}]{MA08}
\bibinfo{author}{\bibfnamefont{J.}~\bibnamefont{Mi\'skiewicz}}
  \bibnamefont{and} \bibinfo{author}{\bibfnamefont{M.}~\bibnamefont{Ausloos}},
  \bibinfo{journal}{Physica A} \textbf{\bibinfo{volume}{387}},
  \bibinfo{pages}{6584} (\bibinfo{year}{2008}).

\bibitem[{\citenamefont{Ausloos and Lambiotte}(2007)}]{5}
\bibinfo{author}{\bibfnamefont{M.}~\bibnamefont{Ausloos}} \bibnamefont{and}
  \bibinfo{author}{\bibfnamefont{R.}~\bibnamefont{Lambiotte}},
  \bibinfo{journal}{Physica A} \textbf{\bibinfo{volume}{382}},
  \bibinfo{pages}{16} (\bibinfo{year}{2007}).

\bibitem[{\citenamefont{Ausloos and Gligor}(2007)}]{6}
\bibinfo{author}{\bibfnamefont{M.}~\bibnamefont{Ausloos}} \bibnamefont{and}
  \bibinfo{author}{\bibfnamefont{M.}~\bibnamefont{Gligor}},
  \bibinfo{journal}{Eur. Phys. J B} \textbf{\bibinfo{volume}{57}},
  \bibinfo{pages}{139} (\bibinfo{year}{2007}).

\bibitem[{\citenamefont{Gligor and Ausloos}(2008{\natexlab{a}})}]{gligor-2008}
\bibinfo{author}{\bibfnamefont{M.}~\bibnamefont{Gligor}} \bibnamefont{and}
  \bibinfo{author}{\bibfnamefont{M.}~\bibnamefont{Ausloos}},
  \bibinfo{journal}{J. Econ. Integration} \textbf{\bibinfo{volume}{23}},
  \bibinfo{pages}{297} (\bibinfo{year}{2008}{\natexlab{a}}).

\bibitem[{\citenamefont{Ausloos and Gligor}(2008)}]{ausloos-2008}
\bibinfo{author}{\bibfnamefont{M.}~\bibnamefont{Ausloos}} \bibnamefont{and}
  \bibinfo{author}{\bibfnamefont{M.}~\bibnamefont{Gligor}},
  \bibinfo{journal}{Acta Phys. Polon. A} \textbf{\bibinfo{volume}{114}},
  \bibinfo{pages}{491} (\bibinfo{year}{2008}).

\bibitem[{\citenamefont{Gligor and Ausloos}(2008{\natexlab{b}})}]{9}
\bibinfo{author}{\bibfnamefont{M.}~\bibnamefont{Gligor}} \bibnamefont{and}
  \bibinfo{author}{\bibfnamefont{M.}~\bibnamefont{Ausloos}},
  \bibinfo{journal}{Eur. Phys. J. B} \textbf{\bibinfo{volume}{63}},
  \bibinfo{pages}{533} (\bibinfo{year}{2008}{\natexlab{b}}).

\bibitem[{\citenamefont{Shalizi and Newman}(2007)}]{q-calc}
\bibinfo{author}{\bibfnamefont{A.~C.~C.} \bibnamefont{Shalizi}}
  \bibnamefont{and} \bibinfo{author}{\bibfnamefont{M.}~\bibnamefont{Newman}},
  \emph{\bibinfo{title}{Power-law distributions in empirical data}},
  \bibinfo{howpublished}{E-print, arXiv:0706.1062} (\bibinfo{year}{2007}).

\bibitem[{\citenamefont{Borges}(2004)}]{qGDP}
\bibinfo{author}{\bibfnamefont{E.~P.} \bibnamefont{Borges}},
  \bibinfo{journal}{Phys. A} \textbf{\bibinfo{volume}{334}},
  \bibinfo{pages}{255} (\bibinfo{year}{2004}).

\end{thebibliography}

\end{document}